\documentclass[10pt,conference,twocolumn]{IEEEtran}

\usepackage{amsmath}
\usepackage{amssymb}
\usepackage{latexsym}
\usepackage{amsfonts}

\usepackage{LATEXCAD}

\begin{document}
\pagestyle{headings}  
\vspace{1cm}

\title{Periodic sequences with stable $k$-error linear complexity}

\author{
\authorblockN{Jianqin Zhou}
\authorblockA{ 1.Telecommunication School, Hangzhou Dianzi University,
Hangzhou, 310018 China\\ 2. Computer Science School, Anhui Univ. of
Technology, Ma'anshan, 243002 China\\ \ \ zhou9@yahoo.com
 }
}
\maketitle              

\begin{abstract}
The linear complexity of a sequence has been used as an important
measure of keystream strength, hence designing a sequence which
possesses high linear complexity and $k$-error linear complexity is
a hot topic in cryptography and communication. Niederreiter first
noticed many periodic sequences with high $k$-error linear
complexity over GF(q). In this paper, the concept of stable
$k$-error linear complexity is presented to study sequences with
high $k$-error linear complexity. By studying linear complexity of
binary sequences with period $2^n$, the method using cube theory to
construct sequences with maximum stable $k$-error linear complexity
is presented.  It is  proved that a binary sequence with period
$2^n$ can be decomposed into some disjoint cubes. The cube theory is
a new tool to  study $k$-error linear complexity. Finally, it is
proved that the maximum  $k$-error linear complexity is
$2^n-(2^l-1)$ over all $2^n$-periodic binary sequences, where
$2^{l-1}\le k<2^{l}$.

\noindent {\bf Keywords:} {\it Periodic sequence; linear complexity;
$k$-error linear complexity; stable $k$-error linear complexity;
cube}

\noindent {\bf MSC2000:} 94A55, 94A60, 11B50
\end{abstract}

\section{Introduction}

The concept of linear complexity is very useful in the study of the
security of stream ciphers for cryptographic applications. A
necessary condition for the security of a key stream generator is
that it produces a sequence with large linear complexity. However,
high linear complexity can not necessarily guarantee the sequence is
safe. The linear complexity of some sequences is unstable. If a
small number of changes to a sequence greatly reduce its linear
complexity, then the resulting key stream is cryptographically weak.
Ding, Xiao and Shan in their book \cite{Ding} noticed this problem
first, and presented the weight complexity and sphere complexity.
Stamp and Martin \cite{Stamp} introduced $k$-error linear
complexity, which is similar to the sphere complexity, and presented
the concept of $k$-error linear complexity profile. Suppose that (s)
is a sequence over GF(q) with period N. For $k(0\le k\le N)$,
$k$-error linear complexity of (s) is defined as the smallest linear
complexity that can be obtained when any $k$ or fewer of the terms
of the sequence are changed within one period. For small $k$,
Niederreiter \cite{Niederreiter} presented sequences over GF(q)
which possess high linear complexity and $k$-error linear
complexity. By generalized discrete Fourier transform, Hu and Feng
\cite{Hu} constructed some periodic sequences over GF(q) which
possess very large 1-error linear complexity.

The reason why people study the stability of linear complexity is
that changing a small number of elements in a sequence may lead to a
sharp decline of its linear complexity. Therefore we really need to
study such sequences, to which even a small number of changes do not
reduce their linear complexity. We introduce the stable $k$-error
linear complexity to describe this problem. Suppose that (s) is a
sequence over GF(q) with period N. For $k(0\le k\le N)$, the
$k$-error linear complexity of (s) is defined as stable when any $k$
or fewer of the terms of the sequence are changed within one period,
the linear complexity does not decline. By studying the linear
complexity of binary sequences with period $2^n$, a method using
cube theory to construct sequences which possess maximum stable
$k$-error linear complexity is presented, and some examples are
given to illustrate the approach.  It is  proved that a binary
sequence with period $2^n$ can be decomposed into some disjoint
cubes. Therefore, the cube theory is a new tool to study $k$-error
linear complexity. Finally, it is proved that the maximum  $k$-error
linear complexity is $2^n-(2^l-1)$ over all $2^n$-periodic binary
sequences, where $2^{l-1}\le k<2^{l}$.

\section{Preliminaries}

We will consider sequences over GF(q), which is the finite field of
order q. Let $x=(x_1,x_2,\cdots,x_n)$ and $y=(y_1,y_2,\cdots,y_n)$
be vectors over GF(q). Then define
$x+y=(x_1+y_1,x_2+y_2,\cdots,x_n+y_n)$.

The generating function of a sequence $s=\{s_0, s_1, s_2, s_3,
\cdots, \}$  is defined by $$s(x)=s_0+ s_1x+ s_2x^2+ s_3x^3+
\cdots=\sum\limits^\infty_{i=0}s_ix^i$$

The generating function of a finite sequence $s^N=\{s_0, s_1, s_2,
 \cdots, s_{N-1},\}$ is defined by $s^N(x)=s_0+ s_1x+ s_2x^2+
\cdots+s_{N-1}x^{N-1}$. If $s$ is a periodic sequence with the first
period $s^N$, then,
\begin{eqnarray}
s(x) &=& s^N(x)(1+ x^N+ x^{2N}+ \cdots)=\frac{s^N(x)}{1-x^N}\notag\\
&=&\frac{s^N(x)/\gcd(s^N(x),1-x^N)}{(1-x^N)/\gcd(s^N(x),1-x^N)}\notag\\
&=&\frac{g(x)}{f_s(x)}\label{formula01}
\end{eqnarray}
where $f_s(x)=(1-x^N)/\gcd(s^N(x),1-x^N),
g(x)=s^N(x)/\gcd(s^N(x),1-x^N)$.

Obviously, $\gcd(g(x),f_s(x))=1, \deg(g(x)<\deg(f_s(x)))$. $f_s(x)$
is called  the minimal polynomial of $s$, and the degree of $f_s(x)$
is called the linear complexity of $s$, that is $\deg(f_s(x))=L(s)$.

Suppose that N=$2^n$, then $1-x^N=1-x^{2^n}=(1-x)^{2^n}=(1-x)^N$.
Thus for binary sequences with period $2^n$, its linear complexity
is equal to the degree of factor $(1-x)$ in $s^N(x)$.

\noindent {\bf Lemma  2.1} Suppose that s is a binary sequence with
period N=$2^n$, then L(s)=N if and only if the Hamming weight of a
period of the sequence is odd.

\begin{proof}\
 As  L(s)=N$-\deg(\gcd(s^N(x),1-x^N))$, thus L(s)=N if and only if $\gcd(s^N(x),1-x^N)=1$.
  There is no factor $(1-x)$ in $s^N(x)$, so $s^N(1)=1$, hence the Hamming weight of a period of the sequence is odd.
\end{proof}\

If an element one is removed from a sequence whose Hamming weight is
odd, the Hamming weight of the sequence will be changed to even, so
the main concern hereinafter is about sequences whose Hamming weight
are even.

\noindent {\bf Lemma 2.2}  Let $s_1$ and $s_2$ be binary sequences
with period N=$2^n$. If $L(s_1)\ne L(s_2)$, then
$L(s_1+s_2)=\max\{L(s_1),L(s_2)\} $; otherwise if $L(s_1)= L(s_2)$,
then $L(s_1+s_2)<L(s_1)$.

\begin{proof}\
 If  $L(s_1)> L(s_2)$, $s_1(x)$ and $s_2(x)$ are  generating functions of the first period of $s_1$ and $s_2$
  respectively, then

 $s_i(x)=(1-x)^{N-L(s_i)}g_i(x)$, $g_i(1)\ne0, i=1,2$.

Thus $s_1(x)+s_2(x)=(1-x)^{N-L(s_1)}g(x), g(1)\ne0$.

It follows that $L(s_1+s_2)=\max\{L(s_1),L(s_2)\}$.

If $L(s_1)= L(s_2)$, then
$s_1(x)+s_2(x)=(1-x)^{N-L(s_1)}(g_1(x)+g_2(x)), g_1(1)+g_2(1)=0$.

Thus $L(s_1+s_2)<L(s_1)$.
\end{proof}\

Suppose that the linear complexity of s can decline when at least
$k$ elements of s are changed. By Lemma 2.2, the linear complexity
of the binary sequence, in which elements at exactly those $k$
positions are all nonzero, must be L(s). Therefore, for the
computation of $k$-error linear complexity, we only need to find the
binary sequence whose Hamming weight is minimum and its linear
complexity is L(s).

\noindent {\bf Lemma  2.3} Suppose that $E_i$ is a $2^n$-periodic
binary sequence  with one nonzero element at position $i$ and 0
elsewhere in each period, $0\le i\le N$. If $j-i=2^r(1+2a), a\ge0,
0\le i<j<N, r\ge0$, then $L(E_i +E_j)=2^n-2^r$.

\begin{proof}\
Let $E_i +E_j$ correspond to a polynomial, which is given by
$$x^i+x^j=x^i(1+x^{j-i})=x^i(1-x^{j-i})=x^i(1-x^{2^r+2a2^r})$$ where
$a$ is a nonnegative integer.

Note that $1-x^{2^r+2a2^r}=(1-x^{2^r})(1+x^{2^r}+x^{2\cdot2^r}
+\cdots+  x^{2a\cdot2^r})=(1-x^{2^r})f(x)$ and $f(1)=1$, thus

$\gcd((1-x)^{2^n},x^i(1-x^{j-i}))=\gcd((1-x)^{2^n},1-x^{2^r})=\gcd((1-x)^{2^n},(1-x)^{2^r})=(1-x)^{2^r}$

Hence, $L(E_i +E_j)=2^n-2^r$.
\end{proof}\

Denote $E_{ij}$ as a binary sequence with period $2^n$, and it has
only 2 nonzero elements in a period. If there are only 2 adjacent
positions with nonzero element  in $E_{ij}$, then its linear
complexity is $2^n-1$, namely $E_{ij}$ is a sequence with even
Hamming weight and the largest linear complexity. According to Lemma
2.2, if sequence s can be decomposed into the sum of several
$E_{ij}$, in which each has linear complexity  $2^n-1$, and the
number of $E_{ij}$ is odd, then L(s) = $2^n-1$. After a symbol of s
is changed, its Hamming weight will be odd, so its linear complexity
will be $2^n$, namely the 1-error linear complexity of sequence s is
$2^n-1$.

\noindent {\bf Theorem  2.1}  If s is a binary sequence with period
$2^n$, then its maximum 1-error linear complexity is $2^n-1$.

In order to discuss the maximal 2-error linear complexity of a
binary sequence with period $ 2^n$, we now consider a binary
sequence which has only 4 positions with nonzero element.

\noindent {\bf Lemma  2.4} If s is a binary sequence with period
N=$2^n$ and there are only four non-zero elements, thus s can be
decomposed into the sum of two $E_{ij}$. Suppose that non-zero
positions of the first $E_{ij}$ are $i$ and $j$, $j-i=2^d$(1+2u),
and non-zero positions of the second $E_{ij}$ are $k$ and $l$,
$l-k=2^e$(1+2v), $i<k, k-i=2c+1$. If $d=e$, the linear complexity is
$2^n-(2^d+1)$, otherwise $2^n-2^{\min(d,e)}$.

\begin{proof}\
According to Lemma 2.2, if $d\ne e$, then L(s)= $2^n-2^{\min(d,e)}$.

Consider the case of d=e. The corresponding polynomial of $E_i+E_j$
is given by
\begin{eqnarray*}
x^i+x^j&=&x^i(1-x^{j-i})=x^i(1-x^{2^d(1+2u)})\\
&=&x^i(1-x^{2^d})(1+x^{2^d}+x^{2\cdot2^d}+\cdots +x^{2u\cdot2^d})
 \end{eqnarray*}

The corresponding polynomial of $E_k+E_l$ is given by
\begin{eqnarray*}
x^k+x^l&=&x^k(1-x^{l-k})=x^k(1-x^{2^d(1+2v)})\\
&=&x^k(1-x^{2^d})(1+x^{2^d}+x^{2\cdot2^d}+\cdots +x^{2v\cdot2^d})
 \end{eqnarray*}

Then $E_i +E_j+E_k +E_l$ corresponds to a polynomial, which is given
by
\begin{eqnarray*}
&&x^i+x^j+x^k+x^l\\
&=&x^i(1-x^{2^d})[(1+x^{2^d}+x^{2\cdot2^d}+\cdots +x^{2u\cdot2^d})\\
&&\ \ \ \ \ +x^{k-i}(1+x^{2^d}+x^{2\cdot2^d}+\cdots +x^{2v\cdot2^d})]\\
&=&x^i(1-x^{2^d})[1+x^{k-i}+(x^{2^d}+x^{2\cdot2^d}+\cdots +x^{2u\cdot2^d})\\
&&\ \ \ \ \ +x^{k-i}(x^{2^d}+x^{2\cdot2^d}+\cdots +x^{2v\cdot2^d})]\\
&=&x^i(1-x^{2^d})[1+x^{2c+1}+(x^{2^d}+x^{2\cdot2^d}+\cdots +x^{2u\cdot2^d})\\
&&\ \ \ \ \ +x^{k-i}(x^{2^d}+x^{2\cdot2^d}+\cdots +x^{2v\cdot2^d})]\\
&=&x^i(1-x^{2^d+1})[(1+x+x^{2}+\cdots+x^{2c})\\
&&\ \ \  +(x^{2^d}+x^{3\cdot2^d}+\cdots +x^{(2u-1)\cdot2^d})(1+x)^{2^d-1}\\
&&\ \ \  +x^{k-i}(x^{2^d}+x^{3\cdot2^d}+\cdots
+x^{(2v-1)\cdot2^d})(1+x)^{2^d-1}]
 \end{eqnarray*}

There is no factor (1+x) in $(1+x+x^{2}+\cdots+x^{2c})$, hence
$\gcd((1-x)^{2^n},x^i+x^j+x^k+x^l)=(1-x)^{2^d+1}$, thus, L(s)=
$2^n-(2^d+1)$.
\end{proof}\

\noindent {\bf Lemma  2.5} If s is a binary sequence with period
$2^n$ and there are only 4 non-zero elements, and s can be
decomposed into the sum of two $E_{ij}$, in which each has linear
complexity $2^n-1$, then the linear complexity of s is $2^n-(2^d+1)$
or $2^n-2^d, d>0$.

\begin{proof}\
 Suppose that non-zero positions of the first $E_{ij}$ are i and j, whose linear complexity is $2^n-1$, $j-i=2a+1$,
 and non-zero positions of the second $E_{ij}$ are k and $l$, whose linear complexity is also $2^n-1$, $i<k, l-k=2b+1$.

 1) $i<k<l<j$, and $k-i=2c$.

As $j-i=2a+1,l-k=2b+1$, so $$j-l=2a+1-(2b+1+2c)=2(a-b-c)$$

If $j-l=2^d+2u2^d, k-i=2^e+2v2^e$, without loss of generality,
assume $d<e$, by Lemma 2.2, L(s)= $2^n-2^d$, $d>0$.

 If d=e, by Lemma 2.4, since
$l-i=2(b+c)+1$, so L(s)= $2^n-(2^d+1)$.

2) $i<k<l<j$, and $k-i=2c+1$.

As $j-i=2a+1,l-k=2b+1$, so $l-i=2b+1+2c+1=2(b+c+1),
j-k=2a+1-(2c+1)=2(a-c)$

 If $j-k=2^d+2u2^d, l-i=2^e+2v2^e,$ without loss of
generality, assume $d<e$, by Lemma 2.2, L(s)= $2^n-2^d$, $d>0$.

Since $k-i=2c+1$, by Lemma 2.4, if d=e, then L(s)= $2^n-(2^d+1)$.

3) $i<k<j<l$, and $k-i=2c$.

As $j-i=2a+1,l-k=2b+1$,so $j-k=2a+1-2c=2(a-c)+1,
l-j=2b+1-[2(a-c)+1]=2(b+c-a)$

If $l-j=2^d+2u2^d, k-i=2^e+2v2^e$, without loss of generality,
assume $d<e$, by Lemma 2.2, L(s)= $2^n-2^d,d>0$.

Since $j-i=2a+1$, by Lemma 2.4, if d=e, then L(s)= $2^n-(2^d+1)$.

4) $i<k<j<l$, and $k-i=2c+1$.

As $j-i=2a+1,l-k=2b+1$,so $j-k=2a+1-(2c+1)=2(a-c),
l-i=2b+1+2c+1=2(b+c+1)$.

 If $l-i=2^d+2u2^d, j-k=2^e+2v2^e$, without loss
of generality, assume $d<e$, by Lemma 2.2,L(s)= $2^n-2^d$,$d>0$.

Since $k-i=2c+1$, by Lemma 2.4, if d=e, then L(s)= $2^n-(2^d+1)$.

5) $i<j<k<l$, and $k-i=2c$.

As $j-i=2a+1,l-k=2b+1$, so $k-j=2c-(2a+1)=2(c-a)-1,
l-j=2b+1+[2(c-a)-1]=2(b+c-a)$

 If $l-j=2^d+2u2^d, k-i=2^e+2v2^e$, without
loss of generality, assume $d<e$, by Lemma 2.2, L(s)= $2^n-2^d,d>0$.

Note that $j-i=2a+1$, by Lemma 2.4, if d=e, then L(s)=
$2^n-(2^d+1)$.

6) $i<j<k<l$, and $k-i=2c+1$.

As $j-i=2a+1,l-k=2b+1$, so $k-j=2c+1-(2a+1)=2(c-a),
l-i=2b+1+2c+1=2(b+c+1)$

 If $l-i=2^d+2u2^d, k-j=2^e+2v2^e$, without loss of
generality, assume $d<e$, by Lemma 2.2, L(s)= $2^n-2^d,d>0$.

Note that $k-i=2c+1$, by Lemma 2.4, if d=e, then L(s)=
$2^n-(2^d+1)$.

Based on 6 cases above, we conclude that the lemma can be
established.
\end{proof}\

\noindent {\bf Corollary   2.1} Suppose that s is a binary sequence
with period $2^n$ and there are only 4 non-zero elements, and s can
be decomposed into the sum of two $E_{ij}$. If non-zero positions of
the first $E_{ij}$ are i and j, $j-i$ is an odd number, and non-zero
positions of the second $E_{ij}$ are k and $l, l-k$ is an odd number
too, and $i<k, k-i=4c+2, |l-j|=4d+2$, or $|k-j|=4c+2, |l-i|=4d+2$,
then the linear complexity is $2^n-3$.

\begin{proof}\
 According to case 1), 3) and 5) of Lemma 2.5, if $k-i=4c+2, |l-j|=4d+2$, then $|l-j|=2+4d, k-i=2+4c$.
  By Lemma 2.4, note that $j-i=2a+1$, so L(s)= $2^n-(2+1)$.

According to case 2), 4) and 6) of Lemma 2.5, if $|k-j|=4c+2,
|l-i|=4d+2$, then it is easy to know that $k-i$ is odd, thus
$|k-j|=2+4c, |l-i|=2+4d$. By Lemma 2.4, L(s)= $2^n-(2+1)$.
\end{proof}\

\noindent {\bf Corollary   2.2} If  s is a binary sequence with
period $2^n$ and there are only 4 non-zero elements, and s can be
decomposed into the sum of two $E_{ij}$, in which each has linear
complexity $2^n-2$, then the linear complexity of s is
$2^n-(2^d+1)2^e,e=0,1, d>0$ or $2^n-2^d, d>1$.

\begin{proof}\
 Suppose that non-zero positions of the first $E_{ij}$ are i and j, $j-i=4a+2$,
 and non-zero positions of the second $E_{ij}$ are k and $l, l-k=4b+2$, where $i<k$.

If $k-i=2c+1$, according to Lemma 2.4, then L(s)= $2^n-(2+1)$.

If $k-i=2c$, the corresponding polynomial of $E_i+E_j +E_k+E_l$ is
given by

$x^i+x^j+x^k+x^l=x^i(1+x^{j-i}+x^{k-i}+x^{l-k+k-i})$

Therefore, we only need to consider

$1+x^{j-i}+x^{k-i}+x^{l-k+k-i}=1+(x^2)^{2a+1}+(x^2)^c+(x^2)^{2b+1+c}=1+y^{2a+1}+y^c+y^{2b+1+c}$

According to Lemma 2.5, L(s)= $2^n-(2^d+1)2, d>0$ or $2^n-2^d, d>1$.
\end{proof}\

It is easy to get the following conclusions according to Lemma 2.5
and Corollary 2.2.

\noindent {\bf Theorem 2.2} Suppose that s is a binary sequence with
period $2^n$ and there are four non-zero elements, then the
necessary and sufficient conditions for the linear complexity of s
being $2^n-3$ are given by: s can be decomposed into the sum of two
$E_{ij}$, in which each has linear complexity $2^n-2$, if non-zero
positions of the first $E_{ij}$ are i and $k, k-i=4c+2$, and
non-zero positions of the second $E_{ij}$ are j and $l, l-j=4d+2$,
where $i<j$, then $j-i=2a+1$(or $|l-k|=2b+1$ or $|l-i|=2e+1$ or
$|k-j|=2f+1$).

 \unitlength=0.04350mm
\begin{picture}(2200,1200)(-80,-50)

\drawcenteredtext{130}{1000}{$k$} \drawcenteredtext{1070}{1000}{$l$}
\drawcenteredtext{600}{1070}{$2b+1$}

\drawcenteredtext{130}{200}{$i$} \drawcenteredtext{1070}{200}{$j$}
\drawcenteredtext{600}{130}{$2a+1$}

\drawpath{200}{1000}{1000}{1000}
\drawpath{1000}{1000}{1000}{200}
\drawcenteredtext{1070}{600}{$4d+2$}
\drawpath{200}{1000}{1000}{200} \drawcenteredtext{400}{800}{$2f+1$}
\drawpath{200}{1000}{200}{200} \drawcenteredtext{100}{600}{$4c+2$}

\drawpath{200}{200}{1000}{200}
\drawpath{1000}{1000}{200}{200} \drawcenteredtext{800}{800}{$2e+1$}
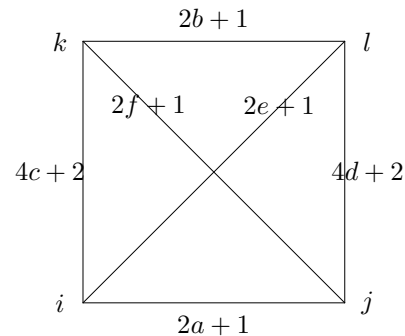
\drawcenteredtext{700.0}{0}{Figure 2.1 A graphic illustration of
Theorem 2.2}
\end{picture}

\noindent {\bf Theorem 2.3} Suppose that s is a binary sequence with
period $2^n$ and its Hamming weight is even, then the maximum stable
2-error linear complexity of s is $2^n-3$.

\begin{proof}\ Assume that L(s) = $2^n-1$, then s can be decomposed into the sum of several $E_{ij}$ and
the number of $E_{ij}$ with linear complexity $2^n-1$ is odd.
According to Lemma 2.2, if an $E_{ij}$ with linear complexity
$2^n-1$ is removed, then the linear complexity of s will be less
than $2^n-1$, namely the 2-error linear complexity of s is less than
$2^n-1$.

Assume that L(s) = $2^n-2$, then s can be decomposed into the sum of
several $E_{ij}$ and the number of $E_{ij}$ with linear complexity
$2^n-2$ is odd. If an $E_{ij}$ with linear complexity $2^n-2$ is
removed, then the linear complexity of s will be less than $2^n-2$,
namely the 2-error linear complexity of s is less than $2^n-2$.

Assume that L(s) = $2^n-3$, without loss of generality, here we only
discuss the case that s has 4 non-zero elements: $e_i, e_j, e_k$ and
$e_l$, and $L(E_i+E_j +E_k+E_l)= 2^n-3$. If any two of them are
removed, by Theorem 2.2, the linear complexity of remaining elements
of the sequence is $2^n-1$ or $2^n-2$. From Figure 2.1, after $e_i$
and $e_l$ are removed, we can see that the linear complexity of the
sequence composed by $e_j$ and $e_k$ is $2^n-1$.

If the position of one element from $e_i, e_j, e_k$ and $e_l$ is
changed, then there exist two elements, of which the position
difference remains unchanged as odd, thus L(s) $\ge 2^n-3$ .

If two nonzero elements are added to the position outside $e_i, e_j,
e_k$ and $e_l$, namely an $E_{ij}$ with linear complexity $2^n-2^d$
is added to sequence s, according to Lemma 2.2, the linear
complexity will be $2^n-1$, $2^n-2$ or $2^n-3$.

The proof is completed.
\end{proof}\

The following is an example to illustrate Theorem 2.3.

 The linear
complexity of 11110$\cdots$0 is $2^n-3$

The linear complexity of 01010$\cdots$0 or 10100$\cdots$0 is $2^n-2$

The linear complexity of 01100$\cdots$0 or 10010$\cdots$0 is $2^n-1$

If two additional nonzero elements are added to 11110$\cdots$0,
namely an $E_{ij}$ whose linear complexity is $2^n-2^d$ is added to
it, according to Lemma 2.2, the linear complexity will become
$2^n-1$, $2^n-2$ or $2^n-3$.

For instance, suppose that 1110$\cdots$010$\cdots$0 is the result of
addition. We only consider that the position difference of the last
two nonzero elements is $2c+1$. According to case 5) of Lemma 2.5,
$j-i=1,l-k=2c+1$, so $k-j=1, l-j=2(c+1)$.

If $l-j=2^d+2u2^d, k-i=2$, according to Lemma 2.2, L(s)= $2^n-2$
when $d>1$.

 If $d=1$, since $j-i=1$, according to Lemma 2.4, L(s)= $2^n-3$.

\section{Cube theory and its main results}

Before presenting some more general results, we first give a special
case.

\noindent {\bf Lemma  3.1} Suppose that s is a binary sequence with
period $2^n$ and there are 8 non-zero elements, thus s can be
decomposed into the sum of 4 $E_{ij}$. Suppose that non-zero
positions of the first $E_{ij}$ are i and j, $j-i=2a+1$, and
non-zero positions of the second $E_{ij}$ are k and $l, l-k=2b+1$,
and $k-i=4c+2, l-j=4d+2$, and non-zero positions of the third
$E_{ij}$ are m and n, non-zero positions of the fourth $E_{ij}$ are
p and q, and $m-i=4+8u, n-j=4+8v, p-k=4+8w, q-l=4+8y$, where
$a,b,c,d,u,v,w$ and $y$ are all non-negative integers, then the
linear complexity of s is $2^n-7$.

\begin{proof}\
According to Corollary 2.1, $L(E_i+E_j+E_k+E_l)=2^n-3$.

It is easy to verify that $E_m+E_n+E_p+E_q$ also satisfies the
conditions of Corollary 2.1, namely its linear complexity is also
$2^n-3$.

Similar to the proof of Lemma 2.4, the corresponding polynomial of
$E_i+E_k+E_m+E_p$ is given by
\begin{eqnarray*}
&&x^i+x^k+x^m+x^p\\
&=&x^i(1-x^{4})[(1+x^{4}+x^{2\cdot4}+\cdots +x^{2u\cdot4})\\
&&\ \ \ \ \ +x^{k-i}(1+x^{4}+x^{2\cdot4}+\cdots +x^{2w\cdot4})]\\
&=&x^i(1-x^{4})[1+x^{k-i}+(x^{4}+x^{2\cdot4}+\cdots +x^{2u\cdot4})\\
&&\ \ \ \ \ +x^{k-i}(x^{4}+x^{2\cdot4}+\cdots +x^{2w\cdot4})]\\
&=&x^i(1-x^{4})[1+x^{4c+2}+(x^{4}+x^{2\cdot4}+\cdots +x^{2u\cdot4})\\
&&\ \ \ \ \ +x^{k-i}(x^{4}+x^{2\cdot4}+\cdots +x^{2w\cdot4})]\\
&=&x^i(1-x^{6})[(1+x^2+x^{4}+\cdots+x^{4c})\\
&&\ \ \  +(x^{4}+x^{3\cdot4}+\cdots +x^{(2u-1)\cdot4})(1+x)^{2}\\
&&\ \ \  +x^{k-i}(x^{4}+x^{3\cdot4}+\cdots
+x^{(2w-1)\cdot4})(1+x)^{2}]
 \end{eqnarray*}

The corresponding polynomial of $E_j+E_l+E_n+E_q$ is given by
\begin{eqnarray*}
&&x^j+x^l+x^n+x^q\\
&=&x^j(1-x^{4})[(1+x^{4}+x^{2\cdot4}+\cdots +x^{2v\cdot4})\\
&&\ \ \ \ \ +x^{l-j}(1+x^{4}+x^{2\cdot4}+\cdots +x^{2y\cdot4})]\\
&=&x^j(1-x^{6})[(1+x^2+x^{4}+\cdots+x^{4d})\\
&&\ \ \  +(x^{4}+x^{3\cdot4}+\cdots +x^{(2v-1)\cdot4})(1+x)^{2}\\
&&\ \ \  +x^{l-j}(x^{4}+x^{3\cdot4}+\cdots
+x^{(2y-1)\cdot4})(1+x)^{2}]
 \end{eqnarray*}

The corresponding polynomial of $E_i+E_j+E_k+E_l+E_m+E_n+E_p+E_q$ is
given by
\begin{eqnarray*}
&&x^i+x^j+x^k+x^l+x^m+x^n+x^p+x^q\\
&=&x^i(1-x^{6})\{(1+x^2+x^{4}+\cdots+x^{4c})\\
&&\ \ \  +(x^{4}+x^{3\cdot4}+\cdots +x^{(2u-1)\cdot4})(1+x)^{2}\\
&&\ \ \  +x^{k-i}(x^{4}+x^{3\cdot4}+\cdots
+x^{(2w-1)\cdot4})(1+x)^{2}\\
&&\ \ \ +x^{j-i}[(1+x^2+x^{4}+\cdots+x^{4d})\\
&&\ \ \  +(x^{4}+x^{3\cdot4}+\cdots +x^{(2v-1)\cdot4})(1+x)^{2}\\
&&\ \ \  +x^{l-j}(x^{4}+x^{3\cdot4}+\cdots
+x^{(2y-1)\cdot4})(1+x)^{2}]\}\\
&=&x^i(1-x^{6})\{1+x^{j-i}+(x^2+x^{4}+\cdots+x^{4c})\\
&&\ \ \  +(x^{4}+x^{3\cdot4}+\cdots +x^{(2u-1)\cdot4})(1+x)^{2}\\
&&\ \ \  +x^{k-i}(x^{4}+x^{3\cdot4}+\cdots
+x^{(2w-1)\cdot4})(1+x)^{2}\\
&&\ \ \ +x^{j-i}[(x^2+x^{4}+\cdots+x^{4d})\\
&&\ \ \  +(x^{4}+x^{3\cdot4}+\cdots +x^{(2v-1)\cdot4})(1+x)^{2}\\
&&\ \ \  +x^{l-j}(x^{4}+x^{3\cdot4}+\cdots
+x^{(2y-1)\cdot4})(1+x)^{2}]\}\\
&=&x^i(1-x^{7})\{1+x+x^{2}+\cdots+x^{2a}\\
&&\ \ \ +x^2(1+x)(1+x^{4}+\cdots+x^{4(c-1)})\\
&&\ \ \  +(x^{4}+x^{3\cdot4}+\cdots +x^{(2u-1)\cdot4})(1+x)\\
&&\ \ \  +x^{k-i}(x^{4}+x^{3\cdot4}+\cdots
+x^{(2w-1)\cdot4})(1+x)\\
&&\ \ \ +x^{j-i}[x^2(1+x)(1+x^{4}+\cdots+x^{4(d-1)})\\
&&\ \ \  +(x^{4}+x^{3\cdot4}+\cdots +x^{(2v-1)\cdot4})(1+x)\\
&&\ \ \  +x^{l-j}(x^{4}+x^{3\cdot4}+\cdots
+x^{(2y-1)\cdot4})(1+x)]\}
 \end{eqnarray*}

 The number of items in $(1+x+x^{2}+\cdots+x^{2a})$  is odd, thus
 $$\gcd((1-x)^{2^n},x^i+x^j+x^k+x^l+x^m+x^n+x^p+x^q)=(1-x)^7$$
\end{proof}\

\unitlength=0.26mm

\begin{picture}(320,320)(18,-5)

\drawcenteredtext{70}{280}{$p$} \drawcenteredtext{45}{250}{$2$}
\drawcenteredtext{160}{280}{$1$}

\drawcenteredtext{250}{280}{$q$} \drawcenteredtext{250}{200}{$4$}
\drawcenteredtext{225}{250}{$2$}

 \drawpath{80}{280}{240}{280}

\drawpath{80}{280}{30}{220}
\drawcenteredtext{20}{220}{$m$} \drawcenteredtext{20}{140}{$4$}

\drawcenteredtext{110}{220}{$1$}

\drawcenteredtext{200}{220}{$n$} \drawcenteredtext{200}{140}{$4$}

\drawpath{30}{220}{190}{220}

\drawpath{190}{220}{240}{280}

\drawpath{30}{220}{240}{280}

\drawcenteredtext{20}{60}{$i$} \drawcenteredtext{200}{60}{$j$}

\drawcenteredtext{110}{60}{$1$}

 \drawpath{30}{60}{190}{60}
\drawpath{30}{60}{30}{220}
\drawpath{190}{60}{190}{220}

\drawcenteredtext{250}{120}{$l$}\drawcenteredtext{225}{170}{$2$}
\drawcenteredtext{225}{90}{$2$}

\drawpath{190}{60}{240}{120}

\drawpath{190}{220}{240}{120}

\drawpath{190}{220}{80}{280}

\drawpath{240}{280}{240}{120}

\drawcenteredtext{70}{120}{$k$} \drawcenteredtext{45}{90}{$2$}
\drawcenteredtext{160}{120}{$1$}


 \drawpath{30.00}{60.00}{35.00}{66.00}
 \drawpath{36.25}{67.50}{40.00}{72.00}
 \drawpath{41.25}{73.50}{45.00}{78.00}
 \drawpath{46.25}{79.50}{50.00}{84.00}
 \drawpath{51.25}{85.50}{55.00}{90.00}
 \drawpath{56.25}{91.50}{60.00}{96.00}
 \drawpath{61.25}{97.50}{65.00}{102.00}
 \drawpath{66.25}{103.50}{70.00}{108.00}
 \drawpath{71.25}{109.50}{75.00}{114.00}
 \drawpath{76.25}{115.50}{80.00}{120.00}


 \drawpath{80.00}{120.00}{96.00}{120.00}
 \drawpath{100.00}{120.00}{112.00}{120.00}
 \drawpath{116.00}{120.00}{128.00}{120.00}
 \drawpath{132.00}{120.00}{144.00}{120.00}
 \drawpath{148.00}{120.00}{160.00}{120.00}
 \drawpath{164.00}{120.00}{176.00}{120.00}
 \drawpath{180.00}{120.00}{192.00}{120.00}
 \drawpath{196.00}{120.00}{208.00}{120.00}
 \drawpath{212.00}{120.00}{224.00}{120.00}
 \drawpath{228.00}{120.00}{240.00}{120.00}


 \drawpath{80.00}{120.00}{80.00}{136.00}
 \drawpath{80.00}{140.00}{80.00}{152.00}
 \drawpath{80.00}{156.00}{80.00}{168.00}
 \drawpath{80.00}{172.00}{80.00}{184.00}
 \drawpath{80.00}{188.00}{80.00}{200.00}
 \drawpath{80.00}{204.00}{80.00}{216.00}
 \drawpath{80.00}{220.00}{80.00}{232.00}
 \drawpath{80.00}{236.00}{80.00}{248.00}
 \drawpath{80.00}{252.00}{80.00}{264.00}
 \drawpath{80.00}{268.00}{80.00}{280.00}


 \drawpath{80.00}{120.00}{91.00}{114.00}
 \drawpath{93.75}{112.50}{102.00}{108.00}
 \drawpath{104.75}{106.50}{113.00}{102.00}
 \drawpath{115.75}{100.50}{124.00}{96.00}
 \drawpath{126.75}{94.50}{135.00}{90.00}
 \drawpath{137.75}{88.50}{146.00}{84.00}
 \drawpath{148.75}{82.50}{157.00}{78.00}
 \drawpath{159.75}{76.50}{168.00}{72.00}
 \drawpath{170.75}{70.50}{179.00}{66.00}
 \drawpath{181.75}{64.50}{190.00}{60.00}


 \drawpath{80.00}{120.00}{75.00}{130.00}
 \drawpath{73.75}{132.50}{70.00}{140.00}
 \drawpath{68.75}{142.50}{65.00}{150.00}
 \drawpath{63.75}{152.50}{60.00}{160.00}
 \drawpath{58.75}{162.50}{55.00}{170.00}
 \drawpath{53.75}{172.50}{50.00}{180.00}
 \drawpath{48.75}{182.50}{45.00}{190.00}
 \drawpath{43.75}{192.50}{40.00}{200.00}
 \drawpath{38.75}{202.50}{35.00}{210.00}
 \drawpath{33.75}{212.50}{30.00}{220.00}



 \drawpath{30.00}{60.00}{51.00}{66.00}
 \drawpath{56.25}{67.50}{72.00}{72.00}
 \drawpath{77.25}{73.50}{93.00}{78.00}
 \drawpath{98.25}{79.50}{114.00}{84.00}
 \drawpath{119.25}{85.50}{135.00}{90.00}
 \drawpath{140.25}{91.50}{156.00}{96.00}
 \drawpath{161.25}{97.50}{177.00}{102.00}
 \drawpath{182.25}{103.50}{198.00}{108.00}
 \drawpath{203.25}{109.50}{219.00}{114.00}
 \drawpath{224.25}{115.50}{240.00}{120.00}

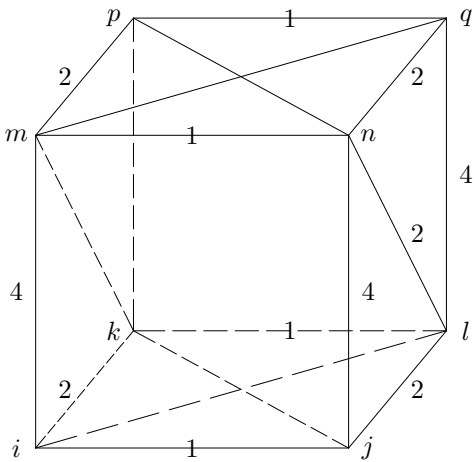
\drawcenteredtext{150.0}{10}{Figure 3.1 A graphic illustration of
Lemma 3.1 }
\end{picture}

For the convenience of presentation, we introduce some definitions.

\noindent {\bf Definition  3.1} Suppose that the difference of
positions of two non-zero elements of sequence s is $(2x+1)2^y$,
both x and y are non-negative integers, then the distance between
the two elements is defined as  $2^y$. If the two elements are the
two ends of an edge, then the length of the edge is defined as
$2^y$.

\noindent {\bf Definition  3.2} Suppose that s is a binary sequence
with period $2^n$, and there are $2^m$ non-zero elements in s, and
$0\le i_1< i_2<\cdots<i_m<n$. If m=1, then there are 2 non-zero
elements in s and the distance between the two elements is
$2^{i_1}$, so it is called as a 1-cube. If m = 2, then s has 4
non-zero elements which form a rectangle, the length of 4 sides are
$2^{i_1}$ and $2^{i_2}$ respectively, so it is called as a 2-cube.
In general, s has $2^{m-1}$ pairs of non-zero elements, in which
there are $2^{m-1}$ non-zero elements which form a (m-1)-cube, the
other $2^{m-1}$ non-zero elements also form a (m-1)-cube, and the
distance between each pair of elements are all  $2^{i_m}$, then the
sequence s is called as an m-cube, and the linear complexity of s is
also called as the linear complexity of the cube.

Similar to the proof of Lemma 3.1, it is easy to prove the following
conclusion.

\noindent {\bf Theorem   3.1} Suppose that s is a binary sequence
with period $2^n$, and non-zero elements of s form a m-cube, length
of  edges are $ i_1, i_2,\cdots ,i_m$ $(0\le i_1< i_2<\cdots<i_m<n
)$ respectively, then L(s)$=2^n-(2^{i_1}+2^{i_2}+\cdots+2^{i_m})$.

There is a 3-cube in Figure3.1. L(s)$=2^n-(1+2+4)$, and length of
edges are $1,2$ and 4 respectively.

\noindent {\bf Theorem   3.2} Suppose that s is a binary sequence
with period $2^n$, and L(s)$=2^n-(2^{i_1}+2^{i_2}+\cdots+2^{i_m})$,
where $0\le i_1< i_2<\cdots<i_m<n$, then the sequence s can be
decomposed into several disjoint cubes, and only one cube has the
linear complexity $2^n-(2^{i_1}+2^{i_2}+\cdots+2^{i_m})$, other
cubes possess distinct linear complexity which are all less than
$2^n-(2^{i_1}+2^{i_2}+\cdots+2^{i_m})$. If the sequence s comprises
only one cube, then the Hamming weight of s is $2^m$.

\begin{proof}\
The mathematical induction will be applied to the degree $d$ of
$s^N(x)$. For $d <3$, by Lemma 2.3, the theorem is established.

A) Suppose that
L(s)$=2^n-(2^{i_1}+2^{i_2}+\cdots+2^{i_m}+2^{i_{m+1}})$, and the
Hamming weight of s is minimum, namely
L(s)$\ne2^n-(2^{i_1}+2^{i_2}+\cdots+2^{i_m}+2^{i_{m+1}})$ when
remove 2 or more non-zero elements. Next we prove that s comprises
one (m+1)-cube exactly. Let

\begin{eqnarray*}
s^N(x)&=&(1-x^{2^{i_1}})(1-x^{2^{i_2}})\cdots(1-x^{2^{i_m}})(1-x^{2^{i_{m+1}}})\\
&&\ \ \ [1+f(x)(1-x)]
\end{eqnarray*}

Then
$t^N(x)=(1-x^{2^{i_1}})(1-x^{2^{i_2}})\cdots(1-x^{2^{i_m}})[1+f(x)(1-x)]$
corresponds to a sequence t whose linear complexity is
L(t)$=2^n-(2^{i_1}+2^{i_2}+\cdots+2^{i_m})$. The degree of $t^N(x)$
is less than the degree of $s^N(x)$, so the mathematical induction
can be applied. In the following, we consider two cases.

1) The Hamming weight of t is $2^m$. By mathematical induction, t is
an m-cube. Since  $s^N(x)=t^N(x)(1-x^{2^{i_{m+1}}})$, and  $0\le
i_1< i_2<\cdots<i_m<i_{m+1}<n$, so s is a (m+1)-cube and its Hamming
weight is $2^{m+1}$.

2)  The Hamming weight of t is $2^m+2y$. By mathematical induction,
$t^N(x)=(1-x^{2^{i_1}})(1-x^{2^{i_2}})\cdots(1-x^{2^{i_m}})[1+g(x)(1-x)+h(x)(1-x)]$,
and
$u^N(x)=(1-x^{2^{i_1}})(1-x^{2^{i_2}})\cdots(1-x^{2^{i_m}})[1+g(x)(1-x)]$,
corresponds to an m-cube, its non-zero elements form a set denoted
by A.

$v^N(x)=(1-x^{2^{i_1}})(1-x^{2^{i_2}})\cdots(1-x^{2^{i_m}})h(x)(1-x)$
corresponds to several cubes, whose 2y non-zero elements form a set
denoted by B.

Assume that  $b\in B, bx^{2^{i_{m+1}}}\in A$, we swap $b$ and
$bx^{2^{i_{m+1}}}$, namely let  $b\in A, bx^{2^{i_{m+1}}}\in B$. It
is easy to show that the linear complexity of the sequence to which
$u^N(x)$ corresponds remains unchanged.

$s^N(x)=t^N(x)(1-x^{2^{i_{m+1}}})=u^N(x)+v^N(x)-u^N(x)x^{2^{i_{m+1}}}-v^N(x)x^{2^{i_{m+1}}}$,
$u^N(x)x^{2^{i_{m+1}}}$ corresponds to $2^m$ non-zero elements which
form a set denoted by C. $v^N(x)x^{2^{i_{m+1}}}$ corresponds to 2y
non-zero elements which form a set denoted by D.

According to Case 1), set A and set C do not have intersection. As
elements of A have low power in the assumption, so set A and set D
do not have intersection.

Set C and B may have intersection, set D and B may have
intersection, but an element e of B can not belong to set C and D
simultaneously.

If $e\in B, e=ax^{2^{i_{m+1}}}\in C, a\in A, e=bx^{2^{i_{m+1}}}\in
D, b\in B$, so a=b. It contradicts the fact that A and B are
disjoint.

Suppose that $b\in B, b=ax^{2^{i_{m+1}}}\in C, a\in A$, then
$ax^{k(2^{i_{m+1}})}(k\ge2)$ must exist in D.

If $k = 2z$ is even, then sequence s has non-zero elements $a$ and
$ax^{2z(2^{i_{m+1}})}$, whose linear complexity is less than
$$2^n-2\cdot2^{i_{m+1}}<2^n-(2^{i_1}+2^{i_2}+\cdots+2^{i_m}+2^{i_{m+1}}).$$
By Lemma 2.2, if the two non-zero elements are removed, the linear
complexity of s remains unchanged. It contradicts the assumption
that the Hamming weight is minimum, so k = 2z +1 is odd.

Thus, the Hamming weight of the sequence to which $s^N(x)$
corresponds is more than or equals to  $|A|+|C|=2^{m+1}$.

$A, C\backslash B$ and $\{ax^{(2k+1)(2^{i_{m+1}})}\in D|a\in A \}$
form a (m+1)-cube exactly, and the linear complexity is
$2^n-(2^{i_1}+2^{i_2}+\cdots+2^{i_m}+2^{i_{m+1}})$.

By the assumption, s has minimum Hamming weight, so 2y non-zero
elements of set B are covered set C or set D, and only the element
$ax^{(2k+1)(2^{i_{m+1}})}(a\in A)$ of set D remains. Namely s
comprises a (m+1)-cube exactly.

B) Let $s^N(x)= u^N(x)+ v^N(x)$, where the Hamming weight of
$u^N(x)$ is minimum, and
$$L(u)=2^n-(2^{i_1}+2^{i_2}+\cdots+2^{i_m}+2^{i_{m+1}}).$$ By Case A),
$u^N(x)$ comprises a (m+1)-cube exactly.

Let $v^N(x)= y^N(x)+ z^N(x)$, where the Hamming weight of $y^N(x)$
is minimum, and L(y)=L(v). By Case A), $y^N(x)$ comprises a cube
exactly. By analogy, we can prove that s comprises several cubes,
and only the linear complexity of one cube is
$2^n-(2^{i_1}+2^{i_2}+\cdots+2^{i_m}+2^{i_{m+1}})$, other cubes
possess distinct linear complexity which are all less than
$2^n-(2^{i_1}+2^{i_2}+\cdots+2^{i_m}+2^{i_{m+1}})$.

The proof is finished.
\end{proof}\

The following examples can help us understand the proof of Theorem
3.2.

$(1+x)(1+x^2)[1+x^5(1+x^2)]=1+x+x^2+x^3+x^5+x^6+x^9+x^{10}$
corresponds a sequence in which there are 8 non-zero elements. It
comprises two 2-cube: $(1+x)(1+x^2)$ and $(1+x)(1+x^4)x^5$.

 $(1+x)(1+x^2)[1+x^5(1+x^2)](1+x^4)=1+x+x^2+x^3+x^4+x^7+x^{13}+x^{14}$ corresponds
a sequence in which there are also 8 non-zero elements, but only one
3-cube. The linear complexity is $2^n-(1+2+4)$, and the length of
edges are 1, 2 and 4 respectively.

Suppose that the linear complexity of s can decline when at least
$k$ elements of s are changed. By Lemma 2.2, the linear complexity
of the binary sequence, in which elements at exactly those $k$
positions are all nonzero, must be L(s). According to Theorem 3.1
and Theorem 3.2, it is easy to get the following conclusion.

\noindent {\bf Corollary   3.1} Suppose that s is a binary sequence
with period  $2^n$, and L(s)$=2^n-(2^{i_1}+2^{i_2}+\cdots+2^{i_m})$,
where  $0\le i_1< i_2<\cdots<i_m<n$. If $k_{\min}$ is the minimum,
such that $k_{\min}$-error linear complexity is less than L(s), then
$k_{\min}=2^m$.

Corollary 3.1 was first proved by Kurosawa et al \cite{Kurosawa},
and later it was proved by Etzion et al \cite{Etzion} in a different
way.

Obviously, previous Theorem 2.2 and Theorem 2.3 are also corollaries
of Theorem 3.1 and Theorem 3.2.

Consider a $k$-cube. The length of edges are 1,2,$2^2,\cdots,$ and
$2^{k-1}$ respectively, and the linear complexity is $2^n-(2^k-1)$.
By Theorem 3.1 and Theorem 3.2, it is easy to get the following
conclusion.

\noindent {\bf Corollary   3.2} Suppose that s is a binary sequence
with period  $2^n$ and its Hamming weight is even, then the maximum
stable $2^{k-1},\cdots, 2^k-2$ or $2^k-1$-error linear complexity of
s are all $2^n-(2^k-1)(k>0)$.

The following is an example to illustrate Corollary 3.2.

Let s be the binary sequence 11$\cdots$110$\cdots$0. Its period is
$2^n$, and there are $2^k$ continuous nonzero elements at the
beginning of the sequence. Then it is a $k$-cube, and the
$2^{k-1},\cdots, 2^k-2$ or $2^k-1$-error linear complexity of s are
all $2^n-(2^k-1)$.

After at most $e(0\le e\le 2^k-1)$ elements of a period in the above
sequence are changed, the linear complexity of all new sequences are
not less than the linear complexity of original sequences, so the
original sequence possesses stable e-error linear complexity.

According to Lemma 2.2, if a sequence whose linear complexity is
less than $2^n-(2^k-1)$ is added to the sequence with linear
complexity $2^n-(2^k-1)$, then the linear complexity of the new
sequence is still $2^n-(2^k-1)$, and the $2^{k-1},\cdots, 2^k-2$ or
$2^k-1$-error linear complexity of the new sequence are all
$2^n-(2^k-1)$.

Combining Corollary   3.1 and Corollary   3.2, it is easy to show
the following theorem.

\noindent {\bf Theorem   3.3} For $ 2^{l-1}\le k<2^{l}$,
$$L_k(s)=\max\limits_tL_k(t)$$ where $s$ is a $2^n$-periodic  binary sequence
with stable $k$-linear complexity $2^n-(2^l-1)$ and $t$ is any
$2^n$-periodic  binary sequence.

CELCS (critical error linear complexity spectrum) is studied  by
Etzion et al \cite{Etzion}. The CELCS of the sequence s comprises
the ordered set of points $(k,c_k(s))$ satisfying $c_k(s)>
c_{k'}(s)$, for $k'>k$; these are the points where a decrease occurs
in the $k$-error linear complexity, and are called critical points.

Let $s$ be a binary sequence whose period is $2^n$ and it has only
one cube. Then s has two critical points.

In the following we study binary sequences which comprise several
cubes. By Theorem 3.2, if s is a binary sequence whose every period
is $2^n$, then it can be decomposed into several cubes. The
following examples show that the cube decomposition of a sequence is
not necessarily unique.

 $1+x+x^3+x^4+x^7+x^8$ can be decomposed into a 1-cube $1+x$, whose linear complexity is $2^n -1$,
  and a 2-cube  $x^3+x^4+x^7+x^8$, whose linear complexity is $2^n -(1+4)$.

It can also be decomposed into a 1-cube $x^3+x^4$, whose linear
complexity is $2^n -1$, a 1-cube $x+x^7$, whose linear complexity is
$2^n -2$, and another 1-cube $1+x^8$, whose linear complexity is
$2^n -8$.

It can also be decomposed into a 1-cube $x^7+x^8$, whose linear
complexity is $2^n -1$, a 1-cube $x+x^3$, whose linear complexity is
$2^n -2$, and another 1-cube $1+x^4$, whose linear complexity is
$2^n -4$.

It can also be decomposed into a 1-cube $1+x^3$, whose linear
complexity is $2^n -1$, a 1-cube $x+x^7$, whose linear complexity
$2^n -2$, and another 1-cube $x^4+x^8$, whose linear complexity is
$2^n -4$.

$\cdots \cdots$

By superposing another sequence over the original one to achieve the
maximal decline of the linear complexity of the new sequence, a
direct method is that the linear complexity of the first cube is
changed to the same as the linear complexity of the second cube.

As an illustrative example, note that the linear complexity of
$x^3+x^4+x^7+x^8$ is $2^n -5$, thus superpose $x^{12}+x^{13}$  over
$1+x+x^3+x^4+x^7+x^8$. As the linear complexity of
$1+x+x^{12}+x^{13}$ is also $2^n -5$, so
$1+x+x^3+x^4+x^7+x^8+x^{12}+x^{13}$ can be decomposed into a 2-cube
$x+x^3+x^7+x^{13}$, whose linear complexity is $2^n-6$, and another
2-cube $1+x^4+x^8+x^{12}$, whose linear complexity is $2^n -12$.

To construct the sequence possessing high stable $k$-error linear
complexity, both the first cube and the second cube should possess
higher linear complexity.

\section{Conclusion}

A small number of element changes may lead to a sharp decline of
linear complexity, so the concept of stable $k$-error linear
complexity has been introduced. By studying the linear complexity of
binary sequences with period $2^n$, especially the linear complexity
will decline when the superposition of two sequences with same
linear complexity, an approach to construct the sequence with stable
$k$-error linear complexity based on cube theory has been derived.
It has been proved that a binary sequence whose period is $2^n$ can
be decomposed into several disjoint cubes, so a new approach to
study $k$-error linear complexity has been given.

Etzion et al \cite{Etzion} proposed to study sequences with two
$k$-error linear complexity value exactly, namely its linear
complexity is only L(s) or 0. So these sequences possess stable
$k$-error linear complexity, but not necessarily maximum stable
$k$-error linear complexity.

By using methods similar to that of the binary sequence, we may
study a sequence with period $p^n$ over $F_p$, where $p$ is a prime
number. The polynomial $1-x^{p^n}=(1-x)^{p^n}$ is over $F_p$. Thus
for a sequence with period $p^n$ over $F_p$, its linear complexity
is equal to the degree of factor $(1-x)$ in $s^N(x)$.

The following are some similar conclusions, whose proof is omitted.

\noindent {\bf Lemma   4.1}   Suppose that s is a sequence with
period $p^n$ over $F_p$. Necessary and sufficient conditions for
$L(s)< p^n$ are: the element sum of one period of the sequence s is
divisible by $p$.

\noindent {\bf Lemma   4.2} Both $s_1$ and $s_2$ are sequences with
period $p^n$ over $F_p$.  If  $L(s_1)\ne L(s_2)$, then
$L(s_1+s_2)=\max\{L(s_1),L(s_2)\}$. If $L(s_1)=L(s_2)$, then
$L(s_1+s_2)\le L(s_1)$.

\noindent {\bf Lemma   4.3}   Suppose that s is a sequence with
period $p^n$ over $F_p$, and  $s^N(x)=ax^k(1-x^l), a\ne0(\mod p)$,
$l=bp^m, b\ne0(\mod p)$, then both the linear complexity and 1-error
linear complexity of sequence s are $p^n-p^m$.

 \section*{ Acknowledgment}
 The research was supported by
Zhejiang Natural Science Foundation(No.Y1100318, R1090138) and NSAF
(No. 10776077).

{\bf Jianqin Zhou} received his B.Sc. degree in mathematics from
East China Normal University, China, in 1983, and M.Sc. degree in
probability and statistics from Fudan University, China, in 1989.
From 1989 to 1999, he was with the Department of Mathematics and
Computer Science, Qufu Normal University, China. From 2000 to 2002,
he worked for a number of IT companies in Japan. From 2003 to 2007,
he was with the Department of Computer Science, Anhui University of
Technology, China.
 From Sep 2006  to
Feb 2007, he was a visiting scholar with the Department of
Information and Computer Science, Keio University, Japan. Since 2008
he has been with the Telecommunication School, Hangzhou Dianzi
University, China

He  published more than 70 papers, and  proved a conjecture posed by
famous mathematician Paul Erd\H{o}s et al. His research interests
include coding theory, cryptography and combinatorics.

\end{document}